\pdfoutput=1
\documentclass[10pt,oneside,pdftex,dvipsnames,a4paper]{article}

\usepackage{etex}
\reserveinserts{18}

\usepackage{a4wide}
\usepackage{setspace,xspace,makeidx,fancyvrb,relsize}
\usepackage{alltt}

\usepackage{amsmath,amsthm,amssymb,amsfonts}
\usepackage{parallel,accents}

\usepackage{amsxtra,amstext,latexsym,dsfont} 

\normalfont
\usepackage{bm}

\usepackage[comma]{natbib}

\usepackage[symbol]{footmisc}
\usepackage{perpage}
\MakePerPage[2]{footnote}

\usepackage[pdftex]{graphicx}
\usepackage[dvipsnames]{xcolor}

\usepackage{pgf,tikz}
\usetikzlibrary{snakes}
\usetikzlibrary{backgrounds}
\usetikzlibrary{arrows}
\usetikzlibrary{patterns}
\usetikzlibrary{fit}
\usetikzlibrary{calc}

\usepackage{lscape} 
\usepackage{shorttoc}

\usepackage{array,longtable,booktabs,float,rotating}
\usepackage{dcolumn,ctable}

\usepackage[labelfont={bf},ruled,labelsep=period,small]{caption}

\usepackage[flushleft]{threeparttable}

\usepackage[linkcolor=black,colorlinks=true,citecolor=black,
bookmarks=true,bookmarksopen=true,bookmarksnumbered=true,bookmarksopenlevel=0,%
hyperindex=true,pdfpagemode=UseOutlines,linktocpage=true,pdfpagelayout=TwoColumnLeft,
pdftitle=Biostatistics,pdfstartview=FitH,pdffitwindow=true,urlcolor=blue]{hyperref}

\usepackage{ragged2e}

\usepackage{multirow,multicol}

\usepackage{fancyhdr,extramarks}

 \makeatletter
 \newcommand\sub{\@startsection%
     {subsubsection}{5}{0mm}{-1\baselineskip}{.01\baselineskip}%
     {\normalfont\itshape}}
 \renewcommand\subsubsection{\@startsection%
     {subsubsection}{3}{0mm}{-1\baselineskip}{.01\baselineskip}%
     {\normalfont\itshape}}
 \makeatother

\makeatletter
        \newcommand\Appendix[2][?]{%
            \refstepcounter{section}%
            \addcontentsline{toc}{appendix}%
                {\protect\numberline{\appendixname~\thesection}#1}%
            {\raggedleft\bfseries \appendixname\
                \thesection\par \centering#2\par}%
                \sectionmark{#1}%
                \@afterheading
                \addvspace{\baselineskip}}
        \newcommand\sAppendix[1]{%
            \raggedleft\bfseries\appendixname\par
            \@afterheading\addvspace{\baselineskip}}
    \makeatother

\normalsize \makeindex

\newcolumntype{A}{>{\centering}p{100pt}}
\newlength\savedwidth

\def\coldot{.}%
{\catcode`\.=\active%
    \gdef.{$\egroup\setbox2=\hbox to \dimen0 \bgroup$\coldot}}
\def\rightdots#1{%
    \setbox0=\hbox{$1$}\dimen0=#1\wd0%
    \setbox0=\hbox{$\coldot$}\advance\dimen0 \wd0%
    \setbox2=\hbox to \dimen0 {}%
    \setbox0=\hbox\bgroup\mathcode`\.="8000 $}
\def\endrightdots{$\hfil\egroup\box0\box2}

\newcolumntype{d}[1]{D{.}{.}{#1}}
\newcolumntype{A}{>{\centering}p{100pt}}
\newcolumntype{.}{D{.}{.}{-1}}
\newcolumntype{P}[2]{>{#1\raggedright\arraybackslash}p{#2}}

\DeclareFontFamily{U}{euc}{}
\DeclareFontShape{U}{euc}{m}{n}{<-6>eurm5<6-8>eurm7<8->eurm10}{}%

\theoremstyle{plain}      
\theoremstyle{plain}      
\theoremstyle{plain}      
\theoremstyle{plain}      
\theoremstyle{definition} 
\theoremstyle{definition} 
\theoremstyle{definition} 
\theoremstyle{plain} 
\theoremstyle{definition} 
\theoremstyle{plain} 
\theoremstyle{definition} 
\theoremstyle{definition} 
\theoremstyle{definition} 

\newcounter{nctr}
\newenvironment{en}{\begin{enumerate}}{\end{enumerate}}

\usepackage[rightbars,color]{changebar}
\cbcolor{blue}

\VerbatimFootnotes

\usepackage{url}

\newcommand\tb{\textbf}
\newcommand\ti{\textit}

\newcommand\mcol{\multicolumn}

\newcommand\bb{\mathbb}

\newcommand\te{\text}
\newcommand\ma[1]{\te{\bf{#1}}}
\newcommand\ca{\mathcal}

\newcommand\op{\operatorname}

\newcommand\as{^\ast}

\newcommand\argmin{\operatornamewithlimits{argmin}}

\newcommand\lt{\left}

\newcommand\pri{^\prime}

\newcommand\qq{\qquad}

\newcommand\rt{\right}

\newcommand\tth{^\text{th}}

\newcommand\R{\bb{R}}  

\newcommand\Bm{\ma{m}} 
\newcommand\bw{\ma{w}} %
\newcommand\bx{\ma{x}}
\newcommand\by{\ma{y}}
\newcommand\bz{\ma{z}}

\newcommand\bJ{\ma{J}} 
\newcommand\bM{\ma{M}} 
\newcommand\bX{\ma{X}}
\newcommand\bY{\ma{Y}}

\newcommand\cI{\ca{I}} 
\newcommand\cM{\ca{M}} 
\newcommand\cP{\ca{P}} 
\newcommand\cX{\ca{X}} %
\newcommand\ga{\gamma}

\newcommand\sig{\sigma}

\newcommand\Ga{\Gamma}

\newcommand\bLa{\bm\Lambda}

\usepackage{color}

\makeatletter
\renewcommand\sub{\@startsection%
    {subsection}{3}{0mm}{-1.5\baselineskip}{.1\baselineskip}%
    {\normalfont\large\itshape}}
\makeatother

\usepackage{fancyhdr}
\begin{document}
\sloppy

\begin{center}
Running Head: \uppercase{Group Analysis of Self-organizing Maps} 
\end{center}
\vspace{3cm}
\begin{center}
\Large{\tb{Group Analysis of Self-organizing Maps based on
           Functional MRI using Restricted Frechet Means}}
\\
\vspace{2.5cm} \normalsize
               Arnaud P.~ Fournel${^{ab}}$, 
               Emanuelle Reynaud${^{b}}$, 
               Michael J.~ Brammer${^{a}}$, \\ 
               Andrew Simmons${^{ac}}$,
               and Cedric E.~ Ginestet${^{ac}}$.
\end{center}
\begin{center}
\vspace{1cm}
  \rm ${^a}$Department of Neuroimaging, Institute of Psychiatry,
  King's College London, UK,
  \rm ${^b}$Laboratoire d'Etude des M\'{e}canismes Cognitifs (EMC),
  EA 3082, Universit\'{e} Lumi\`{e}re Lyon II, France, 
  \rm ${^c}$National Institute of Health Research (NIHR) Biomedical
  Research Centre for Mental Health. \\
\end{center}
\vspace{4.5cm}
\sub{Acknowledgments}
This work was supported by a fellowship and core funds from the UK National Institute
for Health Research (NIHR) Biomedical Research Centre for Mental
Health (BRC-MH) at the South London and Maudsley NHS Foundation Trust
and King's College London. This work has also been funded by
the Guy's and St Thomas' Charitable Foundation as well as the South
London and Maudsley Trustees. APF has received financial support from the Region
Rh\^{o}ne-Alpes and the Universit\'{e} Lumi\`{e}re Lyon 2 through an
Explora’Doc grant. We would also like to thank three anonymous reviewers for their
valuable inputs.

\sub{Correspondence}
Correspondence concerning this article should be sent to Cedric
Ginestet at the Centre for Neuroimaging Sciences, NIHR Biomedical Research Centre,
Institute of Psychiatry, Box P089, King's College London,
De Crespigny Park, London, SE5 8AF, UK. Email may be sent to
\rm cedric.ginestet@kcl.ac.uk
\pagebreak

\begin{abstract}
	Studies of functional MRI data are increasingly concerned with the
        estimation of differences in spatio-temporal networks across
        groups of subjects or experimental conditions. Unsupervised
        clustering and independent component analysis (ICA) have been used
        to identify such spatio-temporal networks. While these
        approaches have been useful for estimating these networks at
        the subject-level, comparisons over groups or experimental conditions
        require further methodological development. In this paper, we
        tackle this problem by showing how self-organizing maps (SOMs) can be compared
        within a Frechean inferential framework. Here, we summarize
        the mean SOM in each group as a Frechet mean with respect to a
        metric on the space of SOMs. The advantage of this approach is
        twofold. Firstly, it allows the visualization of the mean
        SOM in each experimental condition. Secondly, this Frechean
        approach permits one to draw inference on group differences, using
        permutation of the group labels.         
        We consider the use of different distance functions, and introduce two
        extensions of the classical sum of minimum distance (SMD)
        between two SOMs, which take into account the spatio-temporal
        pattern of the fMRI data. 
        The validity of these methods is illustrated on synthetic data. Through
        these simulations, we show that the three distance functions of interest
        behave as expected, in the sense that the ones capturing
        temporal, spatial and spatio-temporal aspects of the SOMs are more
        likely to reach significance under simulated scenarios
        characterized by temporal, spatial and spatio-temporal
        differences, respectively. In addition, a re-analysis of a classical
        experiment on visually-triggered emotions demonstrates the
        usefulness of this methodology. In this study, the multivariate functional patterns
        typical of the subjects exposed to pleasant and unpleasant
        stimuli are found to be more similar than the ones of the
        subjects exposed to emotionally neutral stimuli. In this
        re-analysis, the group-level SOM output units with the
        smallest sample Jaccard indices were compared with standard
        GLM group-specific $z$-score maps, and provided considerable
        levels of agreement. Taken
        together, these results indicate that our proposed methods can
        cast new light on existing data by adopting a global
        analytical perspective on functional MRI paradigms. 
\end{abstract}

\section{Introduction}
Self-organizing Maps (SOMs) were originally introduced by Teuvo
\citet{kohonen2000self}. A SOM is an unsupervised artificial neural
network that describes a training data set as a (typically planar)
layer of neurons or \textit{output units}. Each neuron learns to become a prototype for a
number of input units, until convergence of the algorithm. The resulting SOM therefore represents a
projection of the inputs into a two-dimensional grid. In this sense,
SOMs can be regarded as a dimension-reduction clustering algorithm.
One of the main advantages of this unsupervised
method is that the relative position of the neurons on the grid can
be directly interpreted, in the sense that \textit{proximity} of two
units on the map indicates \textit{similarity} of the prototypal
profiles of these units.

SOMs have proved to be useful for data-driven analysis and have
become popular tools in the machine learning community
\citep{Tarca2007}. In neuroimaging, these methods have been
successfully applied to the detection of fMRI response patterns
related to different cognitive tasks
\citep{Liao:2008ga,Ngan:2002wa,Ngan:1999vd,wismuller2004model}. 
Since SOMs are non-parametric unsupervised neural networks, they do
not require the specification of temporal signal profiles, such as
haemodynamic response function or anatomical regions of
interest in order to generate meaningful summaries of spatio-temporal
patterns of brain activity. As a result, these methods have also been used
to identify variations in low-frequency functional connectivity
\citep{Peltier:2003hp}. Statistically, however, observe that the
absence of a probabilistic model can also be a limitation, as this does not
allow for a formal evaluation of the goodness-of-fit of the
method. 

When used for clustering, the SOMs have the following main
advantages. Firstly, starting with a sufficient number of neurons, the
SOM procedure is able to identify features in the data even when these
features are only typical of a small number of input vectors.
Secondly, the resulting layer of neurons is arranged according to the
similarity of these prototypes in the original data space. This
`topology-preserving' property is generally not available in other
data-reduction techniques, such as independent component
analysis (ICA) or $k$-means clustering. This is one desirable property that SOMs
share with multidimensional scaling. This topological structure facilitates the
merging of nodes in order to form `superclusters', which provide a 
way to visualize and compare high-dimensional fMRI data sets. 
\citet{Fischer1999}, for instance, have demonstrated the specific relevance of
these advantages to the analysis of experimental fMRI data. 

One of the outstanding questions in the application of SOMs to fMRI
data is whether one can summarize several subject-specific SOMs into a
`mean map', which would pool information over several subjects. In
addition, it may be of interest to draw inference over group
differences, by comparing the mean maps of several groups of
subjects. Here, the term \textit{group} is used interchangeably with
the concept of experimental condition. Hence, two distinct groups need not be
composed of different subjects, but may only represent different sets of
measurements on the same individuals.
One may, for instance, be interested in extracting the
SOM that summarizes functional brain activity during a particular
cognitive task; or in comparing the resting-state SOM signature of schizophrenic
patients with that of normal subjects. Few studies, however, have
tackled the problem of formally comparing two or more families of
SOMs. Although several authors have proposed distance functions on
spaces of SOMs \citep{Kaski:1996wr,Deng:2007cw,kirt2007method}, to the
best of our knowledge, none
of these researchers have attempted to draw statistical inference on
the basis of such comparisons. This lack of
methodology highlights a pressing need for developing new strategies that
would permit the extension of such multivariate methods from single-subject
analysis to multiple group comparison. 

SOM group analysis can naturally be articulated within a Frechean statistical
framework. In 1948, Frechet introduced the concept of Frechet mean,
sometimes referred to as barycentre or Karcher mean in the context of
Euclidean and Riemannian geometry, respectively
\citep{Karcher1977}. The Frechet mean extends this concept to any metric space.
This quantity is a generalization of the traditional arithmetic mean, applied to
abstract-valued random variables, defined over a metric space. 
The definition of a generalized notion of the arithmetic mean
therefore solely relies on the specification of a metric on the data
space of interest. Once such a metric has been specified, the Frechet
mean is simply the element that minimizes a convex combination of the
squared distances from all the elements in the space of interest. 
Hence, we can construct a metric space of SOMs by
choosing a metric on that space, which permits the comparison of any two
given SOMs in that space. Note that, in that context, the chosen
pairwise distance function should be a proper metric in the sense that
it should satisfy the four metric axioms: (i) non-negativity, (ii)
coincidence, (iii) symmetry and (iv) the triangle inequality 
\citep[see][for an introduction to metric spaces]{Searcoid2007}. In
the sequel, we consider the use of different distance functions on
spaces of SOMs, which do not satisfy the triangle
inequality. Nonetheless, we will show that such distance functions can
easily be transformed into proper metrics, using a straightforward
manipulation \citep{ThomasEiter:1997ux}. See appendix A.
The concept of the Frechet mean has proved to be useful in several domains
of applications, including image analysis \citep{Thorstensen2009,Bigot2011},
statistical shape analysis \citep{Dryden1998},
and in the study of phylogenetic trees \citep{Balding2009}. 

In this paper, our purpose is to use the concept of the Frechet mean for
drawing statistical inference over several families of subject-specific
SOMs. We thus construct Frechean independent and paired-sample
$t$-statistics, by analogy with the classical treatment of real-valued
random variables. Statistical inference for these different tests are
then drawn using permutation of the group labels. In the paper at hand, these
statistics will be constructed using the \textit{restricted} Frechet
mean, which has been shown to have desirable asymptotic properties
\citep{Sverdrup1981}, but is more convenient to use from a
computational perspective. The restricted Frechet mean is defined as
the element in the sample space, which minimizes the squared distances
from all the elements in the sample. This formal approach to group
inference on families of subject-specific SOMs has the advantage of
allowing a direct representation of the mean SOM in each group,
thereby pooling together subject-specific information. In addition,
the proposed methods also allow to formally draw inference at the
group-level in terms of the chosen distance function. 

The paper is organized as follows. In the next section, we give a
general introduction to SOMs, and how they are computed highlighting
the specific algorithm, which will be used throughout the rest of the
paper. In a third section, we describe our proposed Frechean
framework for drawing inference on several groups of SOMs. This
strategy is entirely reliant on the choice of metric for comparing two
given SOMs, and we therefore dedicate a fourth section to the
description of several distance functions on spaces of SOMs, which appear
particularly well-suited for the analysis of fMRI data. These
methods are tested on synthetic data, under a range of different
conditions in section five, and on a real data set in
section six. We close the paper by discussing the
potential usefulness of this statistical strategy with an emphasis on the
critical importance of the choice of the distance function. 

\section{Self-Organizing Maps (SOMs)}\label{sec:som}
We assume here that an fMRI data set is available, which consists
of several spatio-temporal volumes $\bX_{i}$, with $i=1,\ldots,n_{j}$,
for $n_{j}$ subjects in the $j\tth$ experimental group. Each $\bX_{i}$ is a
$V\times T$ matrix, with $V$ voxels and $T$ time points. In the
sequel, it will be of interest to compare several families of such
volumes, such that $j=1,\ldots,J$, for $J$ experimental
conditions. When describing the SOM inference algorithms, however, we
will focus on a single subject-specific data set, $\bX$. 

\sub{Sequential algorithm}
\label{sub:sequential_algorithm}
A SOM, denoted $\bM$, consists of $K$ output units or \textit{neurons} arranged in a
two-dimensional rectangular grid of size $K$ where $K=k_1\times k_2$. 
For convenience, we here assume that the grids of interest are square
grids, such that $k_{1}=k_{2}$. Thus, the units of a SOM will be
indexed by $k=1,\ldots,K$, where $k$ `reads' the units in SOM
from left to right and top to bottom. Each entry in $\bM$ is hence
denoted by $\Bm_{k}$, and corresponds to the \textit{coordinates} of
that unit in $\bM$. That is, $\Bm_{k}$ is a two-dimensional
vector representing the position of $\Bm_{k}$ in $\bM$, such that, for
instance, $\Bm_{1}=(1,1)$, and $\Bm_{2}=(1,2)$, and so forth.
Each output unit has an associated weight vector $\bw_k$, which
is, in our case, a time series over $T$ data points. 

The sequential SOM algorithm takes a set of $V$ input units, $\bx_{v}$'s,
corresponding to the rows of the input data $\bX$. 
The steps of the procedure will be indexed by 
$\ga=0,\ldots,\Ga$, which denote the iterations of the algorithm, and $\Ga$ is
the final step at which a stopping condition is satisfied. In our
case, the stopping rule is simply the number of iterations, but more
sophisticated convergence-based criteria can be used.
We firstly initialize the output units in $\bM$ as random draws from a
uniform distribution on $\R^{T}$. Secondly, an input
vector, denoted $\bx_{v}$, is randomly chosen amongst the $V$ time series. All
$V$ voxels are selected at each step of the algorithm, and these input
vectors are therefore dependent on $\ga$. We will thus denote this
dependence on the iterations by $\bx_{v}(\ga)$.
For each input vector presented to $\bM$, we
identify the unit in $\bM$, which is the `closest' to the input
$\bx_{v}(\ga)$. Here, closeness is generally measured in terms of Euclidean
distance with respect to the values taken by the input vectors. The
unit in $\bM$, which is the closest to $\bx_{v}(\ga)$ is referred to
as the Best Matching Unit (BMU). The \textit{index} of that BMU, for a given
input vector $\bx_{v}$ at iteration $\ga$, is defined as follows,
\begin{equation}
   c(v,\ga) = \argmin_{k\in \{1,\ldots,K\}} \|\bx_{v}(\ga) - \bw_k\|,
   \label{eq:cga}
\end{equation}
with $\|\cdot\|$ denoting the Euclidean norm on $\R^{T}$. Here, 
$\bx_{v}(\ga)$ and $\bw_{k}$ are $T$-dimensional time series.
Thirdly, we update the BMU and its neighbors. The new values of these
units are defined as a linear relationship of the input vector
$\bx_{v}(\ga)$. For a given $\bx_{v}(\ga)$, the updating rule for the BMU and
its neighbors is the following,
\begin{equation}
   \bw_k(\ga+1) = \bw_k(\ga) + \alpha(\ga)K_{\ga}(\Bm_k,\Bm_{c(v,\ga)})
   \Big(\bx_{v}(\ga) - \bw_k(\ga)\Big),
   \label{eq:learning}
\end{equation}
for every $k=1,\ldots,K$.
After updating their weights, the BMU and its neighbours are closer to
$\bx_{v}(\ga)$ in the sense that they constitute a better representation
of that input vector. These steps are repeated for a fixed number of
iterations, $\Ga$. The updating rule in equation (\ref{eq:learning}) contains two key
parameters: (i) the learning rate, denoted $\alpha(\ga)$ and (ii) the kernel
function represented by $K_{\ga}(\Bm_k,\Bm_{c(v,\ga)})$, which grows
smaller as we consider units in $\bM$, which are further away from the BMU
in the space of coordinates of $\bM$. We describe these two quantities
in turn. 

The learning rate, $\alpha(\ga)$ in equation (\ref{eq:learning}), is a decreasing function of
the number of iterations, $\ga$, which controls the amount of learning
accomplished by the algorithm --that is, the dependence of the values of the units in
$\bM$ on the inputs. By convention, we have $\alpha(\ga)\in[0,1]$ for
every $\ga$. Three common choices for $\alpha(\cdot)$ are a linear function, a
function inversely proportional to the number of iterations and a power
function, such as the following recursive definition,
\begin{equation}   
    \alpha(\ga+1) =
    \left(\frac{\alpha(0)}{\alpha(\ga)}\right)^{\ga/\Ga},
    \label{eq:alpha}
\end{equation}
for every $\ga=1,\ldots,\Ga$. A popular
initialization for the learning rate is $\alpha(0)=0.1$
\citep{Peltier:2003hp,gonzalez2003anomaly}.
Clearly, as the algorithm progresses towards $\Ga$, the
value of $\alpha(\ga)$ decreases towards $0$. 
Note, however, that, in the paper at hand, we use the batch
version of this algorithm, which does not require the specification of
a learning rate. 

In equation (\ref{eq:learning}), we have also made use of the neighborhood kernel,
$K_{\ga}(\Bm_k,\Bm_{c(v,\ga)})$. As for the learning rate, the value taken
by this kernel decreases with the number of iterations, and is
therefore dependent on $\ga$. This dependence on $\ga$ has been
emphasized through a
subscript on $K$. For a given output unit $c(v,\ga)$ in the map $\bM$,
the neighborhood kernel, $K_{\ga}(\Bm_k,\Bm_{c(v,\ga)})$,
quantifies how `close' is $\Bm_{k}$ to the BMU, which has index
$c(v,\ga)$. Observe that this closeness is expressed in terms of
Euclidean distances on the grid \textit{coordinates}. As commonly done
in this field, we here choose a standard Gaussian kernel to formalize
the dependence of each unit on the values of its neighbors, such that 
\begin{equation}
     K_{\ga}(\Bm_k,\Bm_{c(v,\ga)}) =
     \exp\left(-\frac{\|\Bm_k-\Bm_{c(v,\ga)}\|^2}{2\sigma(\ga)^2}\right),
     \label{eq:kernel}
\end{equation}
where $\|\cdot\|^{2}$ represents the two-dimensional dot product. 
Here, $\sigma(\ga)$ is a linear function of the number of iterations,
which controls the size of the neighborhood around the BMU.
This function is defined recursively as $\sigma(\ga+1) = \sigma(0)(1 - \ga/\Ga)$,
where $\sigma(0)$ is a parameter value that represents the initial
neighborhood radius. This parameter is commonly initialized with
respect to the size of the two-dimensional grid, $\bM$, such that
$\sig(0)=k_{1}$, which is the `height' of the output SOM.

\sub{Batch algorithm}\label{sec:batch_algorithm}
A popular alternative to the sequential SOM algorithm described in the
previous section is the \textit{batch SOM algorithm}, which has the
advantage of being more computationally efficient than its sequential
counterpart \citep{vesanto2000}. It has been successfully used
in the context of fMRI analysis \citep{Ngan:2002wa}, in natural
language processing \citep{kohonen2000self}, and
in the face recognition literature \citep{tan2005recognizing}. The main
difference between these two approaches is that the entire training
set is considered at once in the batch SOM algorithm, which permits
the updating of the target SOM with the net effect of all the inputs. 

This `global' updating is performed by replacing the input vector,
denoted $\bx_{v}(\ga)$ in the previous section, with a
weighted average of the input vectors, where the relative weight of
each input vector is proportional to the neighborhood kernel
values. At the $\ga\tth$ step of the algorithm, we are
therefore conducting the following global updating, 
\begin{equation}
  \bw_k(\ga+1) = \frac
    {\sum_{v=1}^V \bx_{v} K_{\ga}(\Bm_k,\Bm_{c(v,\ga)})}
    {\sum_{v=1}^V K_{\ga}(\Bm_k,\Bm_{c(v,\ga)})},
   \label{eq:batch}
\end{equation}
for every $k=1,\ldots,K$. It can easily be seen from equation
(\ref{eq:batch}) that $\bw_{k}(\ga+1)$ is a
convex linear combination of the input vectors, $\bx_{v}$'s, where each
of the $V$ inputs is weighted by $K_{\ga}(\Bm_k,\Bm_{c(v,\ga)})/\sum_{v=1}^V
K_{\ga}(\Bm_k,\Bm_{c(v,\ga)})$, and the sum of these weights is equal
to 1. Another non-negligible advantage of the batch SOM algorithm is that it
removes the dependence of the outputs on the learning rate parameter,
denoted $\alpha(\ga)$, as stated in the previous section. 

Throughout the rest of the paper, we will make use of the batch
algorithm, with $\sig(0)=k_{1}$, and $k_{1}=k_{2}=3$, thereby
producing SOMs of dimensions $3\times 3$. Output units in all SOMs are
initialized randomly. Other groups of researchers in neuroimaging have
used square SOMs \citep{Peltier:2003hp,Liao:2008ga}. However, we have also
investigated using simulated data, whether the specification of
rectangular maps had a significant impact on our proposed inferential
methods (see appendix B).

\section{SOM Group Frechean Inference}\label{sec:frechet}
The question of inferring the statistical significance of the
difference between two families of SOMs can be addressed through the
use of abstract-valued random variables as advocated by
\citet{Frechet1948}. In this approach, random variables are solely
defined with respect to a probability measure on a metric space, $(\cX,d)$
\citep[see][chap.~ 2]{Parthasarathy1967}. Hence, it suffices to
define a metric on the space of interest, in order to obtain a valid
statistical framework. Once such a metric has been chosen, one can
construct the mean element in that space, which is commonly referred
to as the Frechet mean. 

In the paper at hand, we are considering a space of SOMs, which we may
denote by $(\cM,d)$, where $d$ is a metric on that space. A range of
different distance functions for such spaces of SOMs will be described in the next
section. As in most standard fMRI designs, we assume that we have $J$
experimental conditions, with $n_{j}$ subjects in each condition, 
thereby allowing for a different number of subjects in each
experimental condition.
A full data set will be summarized as an array of
SOMs, $\{\bM_{ij}\}$, with $i=1,\ldots,n_{j}$ and $j=1,\ldots,J$, 
such that $\bM_{ij}$ corresponds to the SOM of the $i\tth$
subject in the $j\tth$ condition. Given such a sample of SOMs, we can
then define the Frechet mean for the $j\tth$ condition as follows, 
\begin{equation}
   \widehat{\bM}_j = \argmin_{\bM\pri\in\cM} \frac{1}{(n_j - 1)}
   \sum_{i=1}^{n_j} d(\bM_{ij},\bM\pri)^2,
   \label{eq:frechet mean}
\end{equation}
where we have used the Bessel's correction (i.e. $n_{j}-1$) by analogy with the
real-valued setting. 
Given the complexity of the underlying space of SOMs, such a
minimization may be unwieldy. As a result, it is computationally
more practical to consider the \textit{restricted} Frechet
mean, as introduced by \citet{Sverdrup1981}. The classical Frechet
mean in equation (\ref{eq:frechet mean}) is obtained by identifying
the element in the population of SOMs, which minimizes the average
squared distances from all the elements in the sample. The restricted Frechet
mean, by contrast, is obtained by identifying the element in the
\textit{sample}, which has this property. Hence, the restricted
Frechet mean is computed as follows, 
\begin{equation}
   \overline{\bM}_j= \argmin_{\bM\pri\in\bLa_{j}} \frac{1}{(n_j - 1)}
   \sum_{i=1}^{n_j} d(\bM_{ij},\bM\pri)^2,  
   \label{eq:restricted frechet mean}
\end{equation}
where $\bLa_{j}$ denotes the sampled $n_{j}$ SOMs in the $j\tth$
condition, such that
$\bLa_{j}=\{\bM_{1,j},\ldots,\bM_{n_{j},j}\}$. The restricted Frechet
mean has been shown to be consistent, through a generalization of the
strong law of large numbers due to
\citet{Sverdrup1981}. Asymptotically, $\overline{\bM}_{j}$ converges
almost surely to a subset of the theoretical restricted mean, which takes values
in the support of the target population distribution. In the sequel,
the theoretical restricted Frechet mean for the $j\tth$ condition will be denoted
by $\mu_{j}$, following standard convention. 

Similarly, one can define the condition-specific sample Frechet 
variances. These quantities are simply the
values taken by the criteria, which are minimized in equation
(\ref{eq:restricted frechet mean}), such that the (restricted) Frechet
variance for the $j\tth$ condition is defined with respect to the
restricted Frechet mean in the following manner, for every $j=1,\ldots,J$,
\begin{equation}
    S_{j}^2 = \frac{1}{(n_{j}-1)} \sum_{i=1}^{n_{j}}
    d(\bM_{ij},\overline{\bM}_{j})^{2}.
    \label{eq:frechet variance}
\end{equation}

Using the restricted Frechet mean and variance, it is now possible to
construct a non-parametric $t$-test on the metric space of
SOMs. Here, we therefore assume that we solely have two experimental
conditions, such that $J=2$. The null hypothesis stating that the
(restricted) Frechet means of these two distributions are $\delta_{0}$-separated,
can be formally expressed as follows,
$H_{0}:d(\mu_{1},\mu_{2})=\delta_{0}$. Naturally, our interest will
especially lie in testing the null hypothesis stating that there
is no difference between the theoretical restricted Frechet means,
which corresponds to $H_{0}:d(\mu_{1},\mu_{2})=0$. This can be tested using the
following Frechet $t$-statistic,
\begin{equation}
    t_{F} =\frac{d(\overline{\bM}_{1},\overline{\bM}_{2}) - \delta_{0}}
                {S_{p}\lt(1/n_{1} + 1/n_{2}\rt)^{1/2}},
    \label{eq:t}
\end{equation}
where the denominator, $S_{p}$, is the classical pooled sample variance,
which is defined by analogy with the real-valued setting as
\begin{equation}
    S^{2}_{p} = \frac{(n_{1}-1)S_{1}^{2} +
      (n_{2}-1)S^{2}_{2}}{n_{1}+n_{2}-2}.
    \notag
\end{equation}
In addition, if one is considering two samples of equal sizes and
assuming equal Frechet variances, then the aforementioned
$t_{F}$-statistic for such a mean difference can be defined as follows
\begin{equation}
    t_{F}= \frac{d(\overline{\bM}_{1},\overline{\bM}_{2})}
            {S_{p}/\sqrt{N}},   
\end{equation}
where, in this case, the pooled variance is simply the sum of the
variances of the two samples, such that
$S^{2}_{p}=S^{2}_{1}+S^{2}_{2}$, and with $N=n_{1}+n_{2}$. 
Statistical inference is then conducted using permutation on the group
labels. Although our proposed
$t_{F}$-statistic is a real-valued random variable, its asymptotic
distribution is unknown. Indeed, the behavior of this statistic depends on a large
number of other random variables, which are combined using the
non-linear procedure for obtaining group-level SOMs. As a result,
the permutation-based distribution of $t_{F}$ under the null
hypothesis is not expected to follow a standard $t$-distribution. In
particular, the null distribution of $t_{F}$ need not be symmetric.
Since we are here solely considering a generalization of the
$t$-test, but will be applying this statistic to more than two
experimental conditions, we will also make use of the standard
Bonferroni correction for multiple testing. 

\section{Choice of Distance Functions}\label{sec:metric}
In our proposed approach to group comparison, the choice of the
metric on the space of SOMs is paramount. Different
distance functions capture different aspects of the SOMs under scrutiny. It is
therefore of interest to evaluate group differences with respect to
several choices of distance functions. We here review the main
distance functions, which
have been previously proposed in the literature for comparing two
given SOMs. In addition, we introduce a spatio-temporal sum of minimum
distances, which is especially relevant for the study of fMRI-based
SOMs. 

\sub{Quantization Error and Other Measures}\label{sec:quantization_error}
This measure is not a metric, but a popular tool for evaluating the 
accuracy of the SOM generated from a given data set. The so-called quantization
error measures the average quantization error of the target SOM \citep{kohonen2001self}.
It is defined as the sum of the Euclidean distances between each input
unit, $\bx_v$, and its best matching prototype on $\bM$ --that is, the BMU of $\bx_{v}$.
The quantization error, denoted $Q_{e}$, is thus formally defined as follows,
\begin{equation}
     Q_e(\bM,\bX) = \sum_{v=1}^V \| \bx_v - \bw_{c(v)}\|,
     \notag
\end{equation}
where, as before, $\|\cdot\|$ denotes the $T$-dimensional Euclidean norm
and $c(v)$ is the index of the BMU in $\bM$ with respect
to $\bx_{v}$ as described in equation (\ref{eq:cga}). 
This measure is a good indicator of the convergence
of a SOM, and is often used when assessing the behavior of the
algorithms described in the previous section. In this paper, we will
use a variant of the quantization error in order to identify the
output units, which explain the largest amount of between-subject
`variance' in the data. However, the quantization error does not allow
the computation of the distance between two given SOMs. 

\citet{Kaski:1996wr} have proposed a measure of dissimilarity
between two SOMs. They proceeded by comparing the shortest path on
each SOM after matching a given pair of input vectors. This dissimilarity measure is computed by
comparing the distances between all pairs of data samples on the feature
maps. This method, however, is not computationally efficient, and
would be especially challenging when considering fMRI data sets, where
neuroscientists are commonly handling about 100,000 input vectors --that is, the
voxel-specific time series-- for every subject.

\sub{Sum of Minimum Distances (T-SMD)}
\label{sub:sum_of_minimum_distances}
The Sum of Minimum Distances (SMD) was originally introduced by
\citet{ThomasEiter:1997ux} and has been widely used in
image recognition and retrieval 
\citep{kriegel2004classification,takala2005block,tungaraza2009similarity}.
Moreover, the SMD function and some of its variants have already been
used in order to tackle the problem of comparing several SOMs \citep{Deng:2007cw}.
Given two SOMs, denoted $\bM_x$ and $\bM_y$ for
input data sets $\bX$ and $\bY$, respectively, the SMD can be computed
as follows. For every unit, $\bw_{x}$ in $\bM_x$, we calculate the
Euclidean distance between $\bw_{x}$ and every unit $\bw_y$ in $\bM_y$
in order to retain the unit in $\bM_{y}$ that minimizes this
distance. These minimal distances are summed and then normalized by
the total number of input vectors, denoted $V$, in our case.
This gives an $\bM_x$-to-$\bM_y$ score.
The same procedure is performed in the opposite
direction in order to produce an $\bM_y$-to-$\bM_x$ score. The average of the
$\bM_x$-to-$\bM_y$ and the $\bM_y$-to-$\bM_x$ scores is then
defined as the overall SMD between $\bM_x$ and $\bM_y$. 
Therefore, this distance function compares SOMs on the basis of the dissimilarity of the
time series underlying each output unit. It follows that this
procedure mainly emphasizes temporal differences between the fMRI
volumes of interest. Thus, we will label this classical SMD as
\textit{temporal SMD}, and denote it by T-SMD. It is formally defined as
\begin{equation}
    \op{T-SMD}(\bM_x,\bM_y) = \frac{1}{2V}\Biggl(\sum_{\bw_x \in \bM_x}
        \min_{\bw_y \in \bM_y} d_e(\bw_x,\bw_y) + \sum_{\bw_y \in \bM_y}
        \min_{\bw_x \in \bM_x} d_e(\bw_y,\bw_x)\Biggr),
        \label{eq:smd}
\end{equation}
where $d_{e}(\bw_x,\bw_y) = \|\bw_{x} - \bw_{y}\|$ is the
Euclidean distance between $\bw_{x}$ and $\bw_{y}$ on $\R^{T}$, where
$\bw_{x}$ and $\bw_{y}$ represent $T$-dimensional prototypal time series for
maps $\bM_{x}$ and $\bM_{y}$, respectively. 

It is important to note that the SMD function can be re-written by
treating a map, $\bM_{x}$, as a set of weight vectors, $\bw_{x}$. In
this case, we consider the metric space of all weight vectors,
$\bw_{x}$. This metric space is $(\R^{T},d_{e})$. By a slight abuse of notation,
the SOM, $\bM_{x}$, will be used to denote the set of all output vectors,
$\bw_{x}$ associated with the units in $\bM_{x}$. Therefore, we have
$\bM_{x}\subset \R^{T}$. As a result, we can apply the classical
definition of the distance between the subset of a metric space and
an element of that space, $\bw_{x}\in \R^{T}$, such that
$d(\bw_{x},\bM_{x})=\min\{d(\bw_{x},\bw\pri_{x}):
\bw\pri_{x}\in\bM_{x}\}$. Using these conventions, it becomes possible to reformulate
the SMD function in equation (\ref{eq:smd}) in the following manner as
stated by \citet{ThomasEiter:1997ux},
\begin{equation}
     \op{T-SMD}(\bM_x,\bM_y) = \frac{1}{2V}\Biggl(\sum_{\bw_x \in \bM_x}
        d_e(\bw_x,\bM_y) + \sum_{\bw_y \in \bM_y}
        d_e(\bw_y,\bM_x)\Biggr).
    \notag
\end{equation}

In addition, observe that the SMD function is not in general a proper
metric, in the sense that the triangle inequality may fail to be
satisfied \citep[see][for a
counterexample]{ThomasEiter:1997ux}. However, one can easily produce a
proper metric through the identification of the shortest paths between
any two elements in the space of interest, and then define a new
metric with respect to these shortest paths (see appendix A). It can easily be shown
that such a transformation necessarily produces proper metrics, when
considering metrics based on the SMD function
\citep{ThomasEiter:1997ux}. This particular
procedure can easily be implemented in our case, because we have
focused our attention on the restricted Frechet mean, where the
minimization required to identify the mean element is solely conducted
over the space of the sampled elements. As a result, there exists a
small number of possible shortest paths between every pair of elements
in the sample, which greatly facilitates the required transformation
for producing a proper metric. This procedure was systematically
conducted in the sequel, and therefore all the variants of the SMD
function utilized in this paper are indeed proper metrics. We will
thus assume throughout this paper that all distance functions have been adequately
transformed. We now introduce two novel variants of the SMD function, which take
into account the spatial and spatio-temporal properties of the fMRI
data.

\sub{Spatial SMD}
One may also be interested in quantifying the amount of `spatial overlap'
between two given SOMs. This question is especially pertinent when
analyzing SOMs based on fMRI data sets. Here, we therefore wish to evaluate whether
the units in two different maps correspond to similar subsets of
voxels in the original images. Such a distance can be quantified
through a slight modification of the aforementioned SMD metric, where
the Hamming distance is used in the place of the Euclidean
distance. 
\begin{equation}
  \begin{aligned}
      \op{S-SMD}(\bM_x,\bM_y) = \frac{1}{2V}\Biggl(\sum_{\bw_x \in \bM_x}
      &\min_{\bw_y \in \bM_y} \op{Ham}(S(\bw_{x}),S(\bw_{y})) \\
      &+ \sum_{\bw_y \in \bM_y} \min_{\bw_x \in \bM_x}\op{Ham}(S(\bw_{y}),S(\bw_{x}))\Biggr).
  \end{aligned}
  \label{eq:s-smd}
\end{equation}
with $V$ denoting the number of voxels in the fMRI volumes of interest,
and where $S(\bw_{x})$ denotes the binarized index vector of the voxels, which have been
assigned to unit $\bw_{x}$ in $\bM_{x}$. That is, if the voxel $v$ has
been assigned to $\bw_{x}$, then $S_{v}(\bw_{x})=1$, otherwise, we
have  $S_{v}(\bw_{x})=0$. In addition, we have here
made use of the celebrated Hamming distance, which takes the following
form \citep{Hamming1950}, 
\begin{equation}
      \op{Ham}\Big(S(\bw_x),S(\bw_y)\Big) = \frac{1}{V}\sum_{v=1}^{V}
      \cI\Big\{S_{v}(\bw_{x}) = S_{v}(\bw_{y})\}\Big\},
      \label{eq:hamming}
\end{equation}
where $\cI\{f(x)\}$ stands for the indicator function taking a value
of $1$ if $f(x)$ is true, and $0$ otherwise. 
Here, the term spatial refers to the spatial
distribution of the voxels allocated to a particular output
unit. Hence, S-SMD does not emphasize the spatial location of the
output units, as these allocations are arbitrary, but rather the
spatial distribution of the voxels allocated to that output unit.
Observe that we have here solely considered differences in
the spatial distributions of the best matched pair of SOM units, where
that matching is done through the minimization reported in equation
(\ref{eq:s-smd}). This approach, however,
omits to take into account the similarity of these SOM units as prototypal
time series. Both the spatial and the temporal aspects of these maps
can nonetheless be combined, as described in our proposed
spatio-temporal SMD. 

\sub{Spatio-Temporal SMD}
In this novel variant of the classical SMD, we are quantifying the
amount of spatial overlap between any pair of output units in two distinct
maps. In contrast to the S-SMD described in the previous section,
however, we are here comparing the spatial distributions (i.e. the
sets of voxel indexes assigned to a particular unit) of the units that
are the closest in terms of time series profiles, thereby combining the
temporal and spatial properties of the data. This
spatio-temporal version of the SMD function is defined as follows,
\begin{equation}
      \op{ST-SMD}(\bM_x,\bM_y) = \frac{1}{2}\Biggl(\sum_{\bw_x \in \bM_x}
      \op{Ham}(S(\bw_{x}),S(\bw_{y}\as))
      + \sum_{\bw_y \in \bM_y} \op{Ham}(S(\bw_y),S(\bw\as_{x}))\Biggr),
      \label{eq:st-smd}
\end{equation}
where 
\begin{equation}
    \bw_{y}\as=\argmin_{\bw_{y}\in\bM_{y}}d_{e}(\bw_{x},\bw_{y}),
    \qq\te{and}\qq
    \bw_{x}\as=\argmin_{\bw_{x}\in\bM_{x}}d_{e}(\bw_{y},\bw_{x});
    \notag
\end{equation}
where, again, $d_{e}(\bw_x,\bw_y) = \|\bw_{x} - \bw_{y}\|$ is the
Euclidean distance on $\R^{T}$, and $S(\bw_{x})$ is the index set of the
voxels in $\bX$, whose time
series are best represented by $\bw_{x}$. Albeit the formulae in
equation (\ref{eq:st-smd}) is somewhat convoluted, it corresponds to,
perhaps, the most intuitive perspective on the problem of comparing
SOMs, when using fMRI data. Indeed, we are quantifying the amount of spatial overlap
between units, which are as similar as possible in terms of their
temporal profiles. 

We summarize this section with a concise description of these three
different types of SOM distance functions:
\begin{en}
  \item[i.] \textit{Temporal SMD} (T-SMD) is based on the sum of the minimum
    Euclidean distances between the time series of the output units.
  \item[ii.] \textit{Spatial SMD} (S-SMD) is based the sum of the minimum
    Hamming distances between the sets of voxels allocated to the output units.
  \item[iii.] \textit{Spatio-temporal SMD} (ST-SMD) is based on the sum of
    Hamming distances between the sets of voxels allocated to the
    output units, which are the most similar in terms of their
    time series. 
\end{en}

\section{Synthetic Data Simulations}\label{sec:sim}
The proposed methods were tested on three different simulated
scenarios, with varying degrees of difficulty. 
In particular, by isolating different types of differences between the
two groups of interest, each of these scenarios emphasizes the need
for the use of specific metrics, capturing different aspects of the
spatio-temporal process under investigation. In this sense, our
simulations strive to produce realistic differences between the two 
families of fMRI volumes, where group differences may be either spatial,
temporal or both spatial and temporal. 

\begin{figure}[t]
    \centering
    \includegraphics[width=15cm]{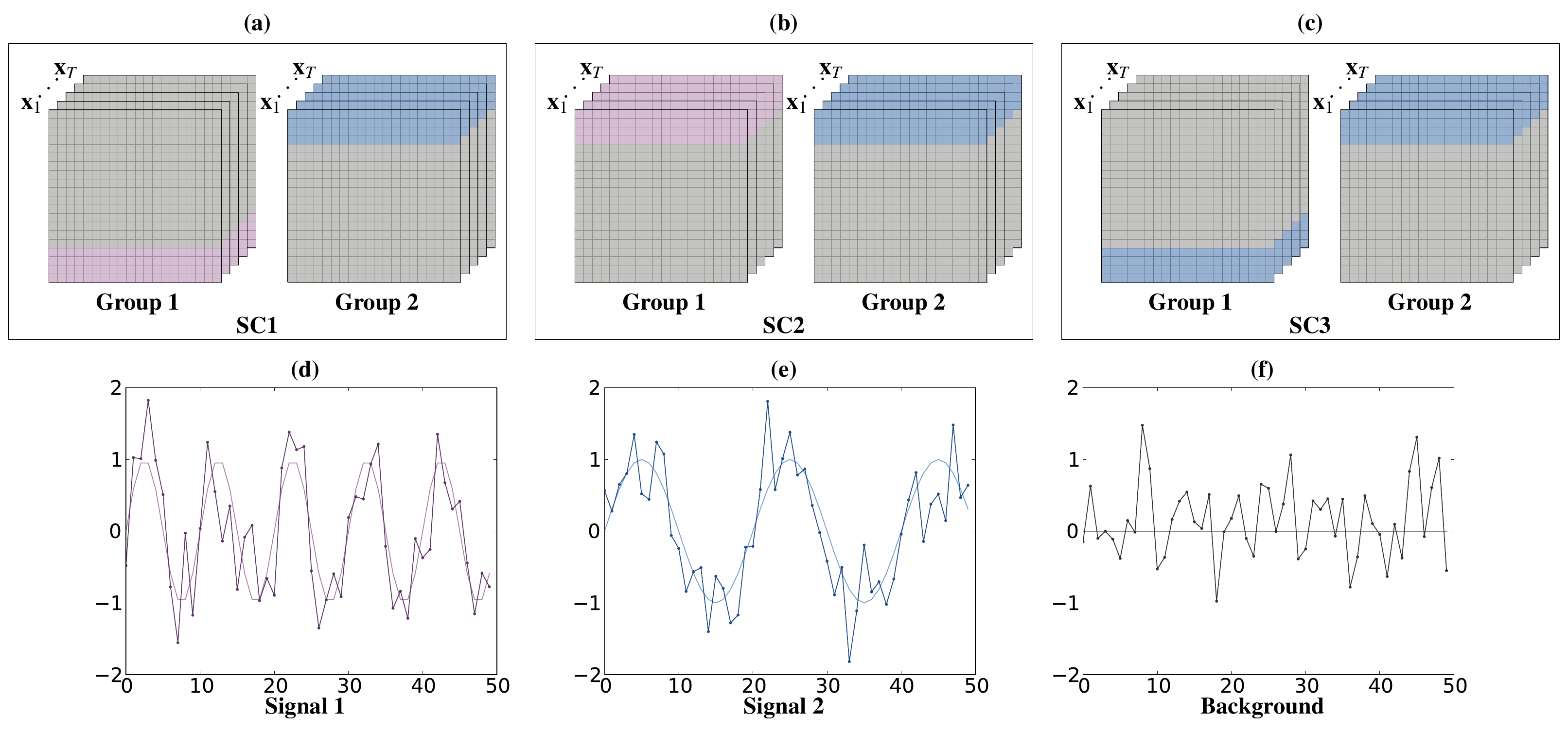}
    \caption{Description of the three simulated scenarios
             ordered by increasing levels of
             difficulty. In panels (a-c), we have reported the
             spatial distributions of the input vectors for each
             scenario, where each data set is composed of 
             $(10\times10)$-images over $T=50$ time points.
             In panels (d-f), we have represented the three types of time
             series used in these simulations. Here, SC1, SC2 and SC3 correspond to
             the three different scenarios, where the two groups exhibit spatio-temporal
             differences (SC1), temporal differences (SC2), and
             spatial differences (SC3), respectively.
             \label{fig:sim}}
\end{figure}
\sub{Simulation Scenarios}
For each simulated data set we have constructed two groups of 
20 subjects, where each subject-specific data set is
composed of two-dimensional images with $10\times 10$ voxels, over 50
time points, as represented in panels (a-c) of figure
\ref{fig:sim}. The time series at each voxel can be of
three different types, as illustrated in panels (d-f) of figure
\ref{fig:sim}, composed of two different signals and one
background time series. The two signals represented in panels (d) and (e) are sinusoids
oscillating over $[-1,1]$, with a frequency of $1/10$Hz and $1/20$Hz,
respectively. We have then added a vector of Gaussian random noise,
$\bz$, to these two types of time series, such that
$z_{t}\sim N(0,\sigma^{2})$, for every $t=1,\ldots,50$,
and for different choices of $\sigma$. The background noise time
series, in panel (f), is solely composed of the random noise for a
given $\sigma$.

The three scenarios in panels (a-c) of figure \ref{fig:sim} are
ordered in terms of the degree of `separability' of the two groups,
where the easiest scenario is on the left and the most difficult one
is on the right. The first data set (SC1) was built with three different
time series at three different locations, corresponding to the
purple, blue and gray colors. In this scenario, the groups differ both in terms of the
temporal profiles of some of their voxels and in terms of the spatial locations
of these voxels. The second scenario (SC2) was constructed with two
different types of time series. Here, the two groups solely
differ in terms of the temporal profiles of some of their
voxels. Finally, in the third scenario (SC3), the two groups only differ
in terms of the spatial locations of the voxels, which have been
assigned the second temporal profile.

In addition, we also studied the effect of the signal-to-noise ratio
(SNR) on the performance of our inferential methods. In particular, we
varied our choice of $\sigma$, when generating the different time
series displayed in panels (d-f) of figure \ref{fig:sim}, in order to
produce different SNRs. In these simulations, the `signal' of interest
was defined as the amplitude of the original sinusoids, which
oscillated between -1 and 1, thus giving an amplitude of
$\lambda=2$. Since the noise affecting this signal was specified to be
Gaussian, the SNR was defined, in our case, as $\op{SNR}=\lambda/2\sig$. Thus, by
setting $\sigma$ to either $1/2$, $1$ or $2$, we produced three
different SNRs of $2$, $1$ and $1/2$, respectively.

These synthetic data sets were analyzed using our proposed inferential
framework. For each simulated subject-specific volume, a SOM 
was computed, and the restricted Frechet mean was identified for each
group. In all scenarios, SOMs were produced by using the batch SOM algorithm.
The output grid was of size $3\times 3$ with $K=9$; 
the number of iterations was set to $100$ steps; 
and we used a decreasing neighborhood kernel of size $k_1=3$, as
commonly done in this field \citep{kohonen2000self}.
For computational convenience, statistical inference was drawn in each
scenario after 100 permutations. Each simulated scenario was
reproduced 100 times, and constructed for the three different levels
of SNR, thereby totalling $900$ distinct simulations.

\begin{table}[t]
\centering
\begin{tabular}{ c | c | c | c | c }
\multicolumn{2}{c|}{\textit{Scenarios and Factors}} & T-SMD & S-SMD & ST-SMD\\
\hline
SC1 (Spatio-temporal) & $\op{SNR} = 2$ & $0\pm0$ & $0.012\pm0.033$ & $0\pm0$ \\
       & $\op{SNR} = 1$ & $0\pm0.001$ & $0.518\pm0.171$ &  $0.003\pm0.012$ \\
       & $\op{SNR} = 0.5$ & $0.030\pm0.066$ & $0.800\pm0.235$ & $0.049\pm0.123$ \\
\hline
SC2 (Temporal) & $\op{SNR} = 2$ & $0\pm0$ & $0.499\pm0.303$ & $0\pm0.006$ \\
       & $\op{SNR} = 1$ & $0\pm0$ & $0.499\pm0.295$ & $0.001\pm0.005$ \\
       & $\op{SNR} = 0.5$ & $0.017\pm0.070$ & $0.484\pm0.296$ & $0.022\pm0.030$ \\
\hline
SC3 (Spatial)& $\op{SNR} = 2$ & $0.472\pm0.294$ & $0.014\pm0.057$ & $0.029\pm0.055$ \\
       & $\op{SNR} = 1$ & $0.464\pm0.286$ & $0.525\pm0.167$ & $0.109\pm0.122$ \\
       & $\op{SNR} = 0.5$ & $0.525\pm0.279$ & $0.783\pm0.271$ & $0.101\pm0.141$ \\
\end{tabular}
\caption{Significance levels based on synthetic data
         with 100 simulations in every cell, with the mean $p$-value
         and the standard deviation for that distribution of
         $p$-values. These results are reported for the three
         scenarios described in figure \ref{fig:sim}, which are denoted by
         SC1, SC2 and SC3, for three different levels of SNR, and for
         the three different distance functions under scrutiny, denoted by
         T-SMD, S-SMD and ST-SMD, which stand for the temporal SMD,
         spatial SMD, and spatio-temporal SMD, respectively. 
         \label{tab:sim}}
\end{table}

\begin{figure}[htbp]
    \centering
    \includegraphics[width=14cm]{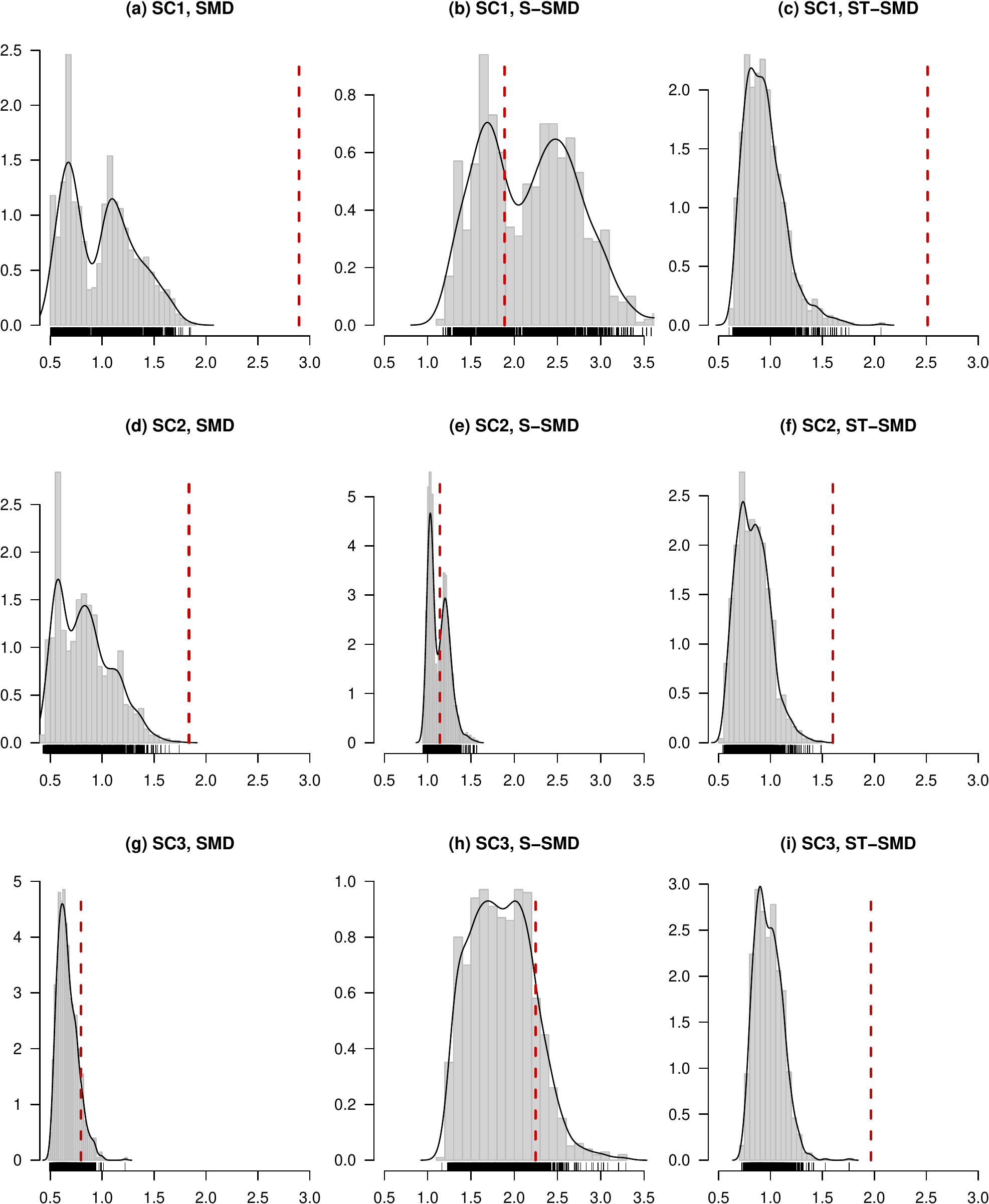}
    \caption{Histograms of the null distributions of 
      $t_{F}$-values obtained through permutation. These null
      distributions are given for a single synthetic data set under
      the three different simulation scenarios, denoted SC1, SC2 and SC3,
      respectively, and for three different metrics on the space of
      SOMs, denoted T-SMD, S-SMD and ST-SMD, which stand for sums of
      minimum distances, spatial T-SMD and spatio-temporal T-SMD,
      respectively. The red dashed line indicates the value of the 
      actual $t_{F}$-statistic for the simulation of interest. These 
      histograms were constructed using data based on an
      $\op{SNR}$ of 1, and for $15$ subjects in each group.
      \label{fig:histos}}
\end{figure}

\sub{Simulation Results}
The summary results of the analysis of these synthetic data sets are reported in
table \ref{tab:sim} and figure \ref{fig:histos}. Overall, the
different metrics of interest were found to successfully capture the
aspects of the simulated SOMs that they were expected to
identify. That is, in the first column of table \ref{tab:sim}, one
can see that T-SMD, which solely takes into account the differences in
voxel-specific temporal profiles, attains its most significant
values under the temporal scenario, SC2. Similarly, in the second
column of table \ref{tab:sim}, the spatial version of the SMD metric,
denoted S-SMD exhibits its best performance under the first and
third scenarios, denoted SC1 and SC3, respectively. Indeed, these two
scenarios are the only ones, where the two groups can be discriminated
in terms of the spatial locations of the different types of time
series. Finally, in the third column of table \ref{tab:sim}, the
spatio-temporal metric, ST-SMD, appears to be optimal under the first
scenario, where group differences can be characterized through
the spatio-temporal properties of the simulated images. 

In addition, we have also evaluated the effect of sample size on
the capacity of the metrics to detect group differences. These results
are not reported in this paper, but we have observed, as expected, that the
statistical power of all the studied metrics improved as the
number of subjects in each group increases. In particular, it was noted that for
the ST-SMD, we solely needed $n\geq 15$ in each group to identify
significant differences under SNR=1, and group
sizes of $n\geq 20$ under SNR=0.5; for all scenarios. Exemplary
null distributions for $t_{F}$-statistics in the three different
scenarios and the three different levels of SNR with $n=15$
are reported in figure \ref{fig:histos}.

In sum, one can note that these three distance
functions exhibit different levels of robustness. In particular, S-SMD
appeared to be especially sensitive to
noise. Although S-SMD succeeded
to capture the spatial differences simulated in SC3, it only
outperformed ST-SMD for high SNRs. Also, in the spatio-temporal
scenario (SC1), T-SMD behaved as well or better than ST-SMD. This
suggests that the T-SMD function is more `powerful'
than the ST-SMD, even when the group differences are characterized by
both spatial and temporal properties. However, the use of ST-SMD
remains justified, because it also succeeds to capture spatial
differences, whereas T-SMD fails to do so. In
general, we therefore recommend the joint use of T-SMD and ST-SMD: If
only T-SMD indicates the presence of group differences, then one can
conclude that such differences are mainly of a temporal nature;
whereas when the use of ST-SMD indicates greater group
differences, this suggests that such differences also have a
spatial character. Overall, these simulations highlight the importance of using several
types of distance functions, as there may not exist a single type of metric,
which would capture all of the aspects of the data of interest. 

\section{Experimental Data}\label{sec:data}
We also evaluated our methods with the re-analysis of a classical
data set, originally published by \citet{Miranda2006}. Since this
first publication, this data set has been re-analyzed several times
with different machine learning algorithms, as conducted by
\citet{Miranda2007} using spatio-temporal support vector machine
(SVM), and \citet{Hardoon2007} with unsupervised methods.

\sub{Subjects and Experimental Design}\label{sec:subjects}
This data set consists of fMRI data from 16 right-handed males with a mean age of 23
years. All participants had normal eyesight and no history of
neurological or psychiatric disorders, and gave written
informed consent to participate in the study, in accordance with the
local ethics committee of the University of North Carolina \citep[see][]{Miranda2006}.
Data were acquired using an experimental block design, composed of three different
conditions: (i) exposure to unpleasant visual stimuli (i.e.~ photos of
dermatological diseases), (ii) exposure to neutral visual stimuli
(i.e.~ photos of neutral day-to-day scenes including human actors) 
and (iii) exposure to male-relevant pleasant visual stimuli (i.e. scantly dressed
women or women in swimsuits). The entire experimental design
consisted of six blocks, where each block contained seven images,
which were each presented to the subjects for three seconds. Each
block was followed by a resting block period where subjects were solely exposed
to a fixation cross. All blocks were of 21$s$ in length. 
\begin{table}[t]
\centering
\begin{tabular}{ l | l | c}
\textit{Metrics} & \textit{Tests} & $p$-values  \\
\hline
T-SMD  & Pleasant vs. Neutral & 0.0\\
       & Unpleasant vs. Neutral & 0.0\\
       & Pleasant vs. Unpleasant & 0.258\\
\hline
S-SMD  & Pleasant vs. Neutral & 0.412\\
       & Unpleasant vs. Neutral & 0.518\\
       & Pleasant vs. Unpleasant & 0.423\\
\hline
ST-SMD & Pleasant vs. Neutral & 0.021\\
       & Unpleasant vs. Neutral & 0.013\\
       & Pleasant vs. Unpleasant & 0.128\\
\end{tabular}
\caption{Significance results of all pairwise comparisons for the three
         conditions of interest, where subjects were exposed to
         pleasant, unpleasant and neutral stimuli. These results are
         reported independently for three different metrics, denoted
         T-SMD, and S-SMD and ST-SMD, which stand for temporal SMD, spatial SMD
         and spatio-temporal SMD, respectively. Observe that since three
         tests were conducted for each pair of conditions and we
         therefore necessitate
         the application of a Bonferroni correction for testing for
         these three pairwise comparisons. Hence, only $p$-values below
         $0.016$ should be regarded as statistically significant. 
         \label{tab:real}}
\end{table}

\sub{Data Acquisition and Pre-Processing}\label{sec:pre-processing}
Blood-oxygenation-level-dependent (BOLD) signal was measured using
a 3-Tesla Allegra head-only MRI System at the Magnetic Resonance Imaging
Research Center in the University of North Carolina. The scanning parameters
were specified as follows. Voxel size was $3\times 3\times 3mm^3$, TR was $3s$,
TE was $30ms$, FA was $80$, FOV was $192\times 192mm$ and each MRI volume
had dimensions $64 \times 64 \times 49$. In each experiment, a total
of $254$ functional volumes were acquired for each participant. 

Data were pre-processed using the FSL Software suite
\citep{Smith:2004eq}; through the use of the Nipype Python Library
\citep{gorgolewski2011}. All fMRI volumes were first motion-corrected
and the skulls were removed, after tissue segmentation. The voxel time series were detrended and
filtered in time and space: that is, low-frequency (drift)
fluctuations were reduced using a band-pass
temporal filter comprised between 0.008Hz and 0.1Hz. The use of a such
a band-pass filter is typical of resting-state connectivity analysis \citep{cordes2001}.
In addition, spatial smoothing was performed using an 8mm
full-width at half-maximum Gaussian kernel.

The first two volumes of each block were discarded, to allow for the
between-block lag in haemodynamic response. The remaining volumes were
concatenated to form three distinct time series representing the three
different conditions. Time series concatenation in the context
of functional connectivity has been introduced by \citet{Fair:2007tv} and
has been implemented by several authors for the study of functional
MRI networks \citep{Ginestet2011a,Ginestet2011b}.
\begin{table}[t]
\centering
\begin{tabular}{ c | c | c | c |}
& \mcol{3}{c}{\textit{Sample Jaccard Index}} \\
\textit{Ranked SOM Units} & \textit{Neutral} & \textit{Pleasant}
& \textit{Unpleasant} \\
\hline
Units up to 1 & \textbf{0.68} & \textbf{0.32} & \textbf{0.55}\\
Units up to 2 & \textbf{0.11} & \textbf{0.28} & \textbf{0.16}\\
Units up to 3 & \textbf{0.08} & \textbf{0.04} & \textbf{0.03}\\
Units up to 4 & 0.01 & 0.07 & 0.07\\
Units up to 5 & 0.03 & 0.08 & 0.03\\
Units up to 6 & 0.01 & 0.05 & 0.04\\
Units up to 7 & 0.05 & 0.08 & 0.05\\
Units up to 8 & 0.01 & 0.02 & 0.05\\
Units up to 9 & 0.02 & 0.06 & 0.02\\
\end{tabular}
\caption{Individual percentage overlaps between the group-level SOM
  components and thresholded GLM $z$-score maps, where the SOM
  components have been ranked with respect to the sample Jaccard
  index. In bold, we have highlighted the three `best'
  condition-specific SOM output units based on T-SMD, which are visually described in
  figures \ref{fig:neutral} to \ref{fig:unpleasant}. Each column sums
  to 1.0, since the concatenated SOM output units cover all the voxels
  in the fMRI volumes of interest.
  \label{tab:comparison}}
\end{table}

\sub{Results of Real-data Analysis}
The significance results of all the pairwise comparisons are reported
in table \ref{tab:real}, after applying the Bonferroni correction for
multiple testing. Our re-analysis of this data set has highlighted a
substantial degree of difference between the neutral and pleasant
conditions, on one hand, and between the neutral and unpleasant
conditions, on the other hand. These pairwise differences were found
to be highly significant under the T-SMD distance function. 
The mean SOM in the unpleasant condition was also found to be
significantly different from the mean SOM in the neutral condition
with respect to the ST-SMD metric ($p<0.016$), albeit to a lesser extent than
under the T-SMD function. The difference between the mean SOMs in the
pleasant condition and the neutral one was found to be just above
significance level under the ST-SMD metric ($p=0.063$). By contrast, 
none of these differences reached significance under the S-SMD,
indicating that differences in spatial allocation of the different
output units of the SOMs in these experimental conditions were not
sufficient to distinguish between the group-level SOMs. 
Importantly, the mean SOMs under the pleasant and unpleasant
conditions were not found to be significantly different under all
three metrics. 

As noted in the analysis of the synthetic data, the fact
that the SOMs are computed on the basis of the similarities between
the voxel-specific time series is likely to be responsible for the
important differences that we are reporting between the metrics
capturing the temporal aspects of the process and
S-SMD, which does not emphasize differences in temporal profiles.
Intriguingly, it should also be noted that the differences between the
SOMs in the pleasant and unpleasant conditions under ST-SMD is lower
than under the T-SMD, thereby perhaps suggesting that the
differences between these two conditions is characterized by an
`interaction' between the temporal and spatial properties of these
functional patterns. 
\begin{figure}[t]
    \centering
    \includegraphics[width=15cm]{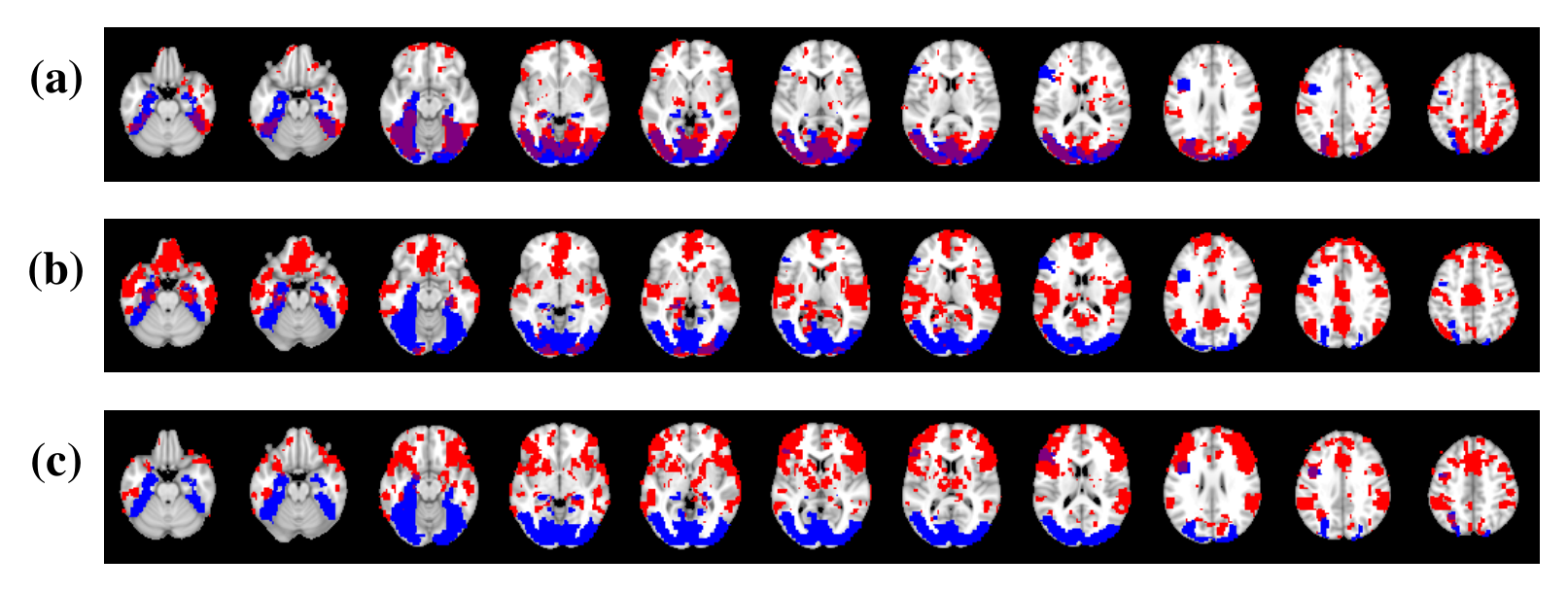}
    \caption{Representation of three output units of the restricted
             Frechet mean SOM for the \textit{unpleasant} condition in
             red, with thresholded ($p\leq0.05$) GLM $z$-score maps in blue. The output
             units have been projected in MNI-normalized space. These
             three output units are the ones that explain the largest
             amount of sample Jaccard index in that SOM. They are
             ordered by Jaccard index from panels (a) to (c),
             with the unit exhibiting the smallest Jaccard index
             in (a).
             \label{fig:neutral}}
\end{figure}
\begin{figure}[t]
    \centering
    \includegraphics[width=15cm]{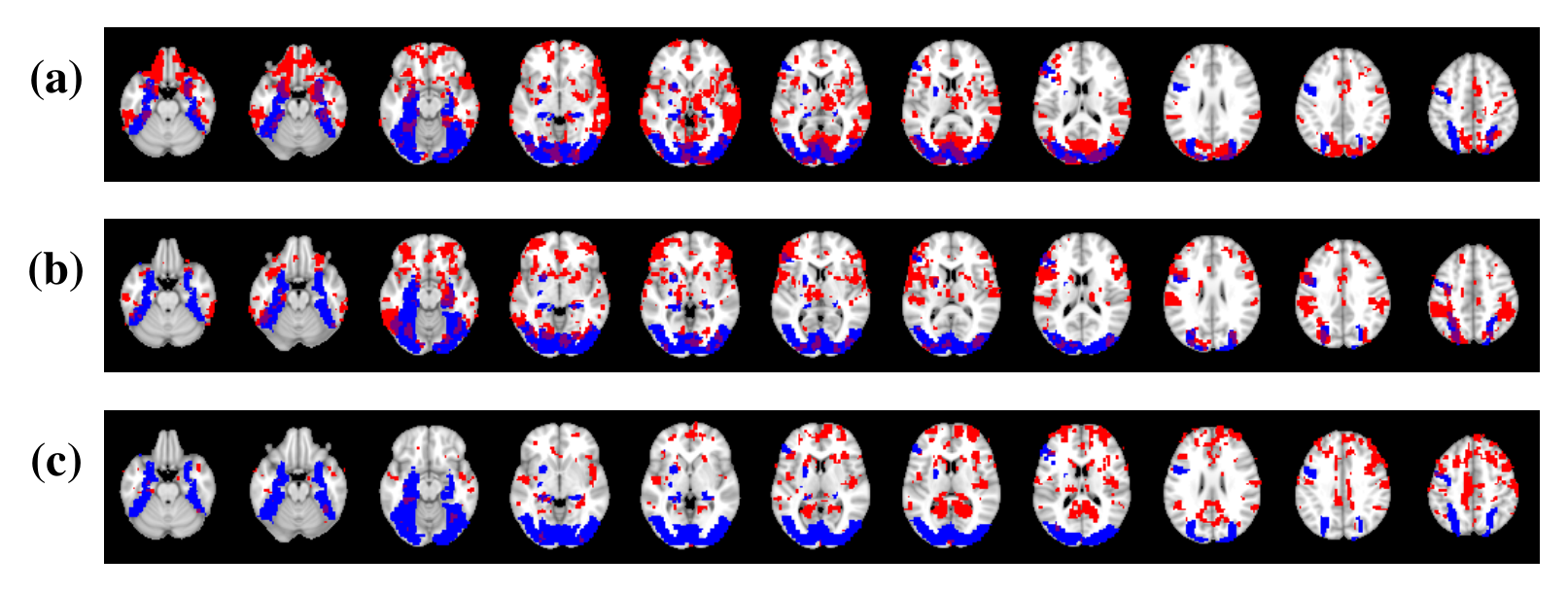}
    \caption{Representation of three output units of the restricted
             Frechet mean SOM for the \textit{unpleasant} condition in
             red, with thresholded ($p\leq0.05$) GLM $z$-score maps in blue. The output
             units have been projected in MNI-normalized space. These
             three output units are the ones that explain the largest
             amount of sample Jaccard index in that SOM. They are
             ordered by Jaccard index from panels (a) to (c),
             with the unit exhibiting the smallest Jaccard index
             in (a).
             \label{fig:pleasant}}
\end{figure}
\begin{figure}[t]
    \centering
    \includegraphics[width=15cm]{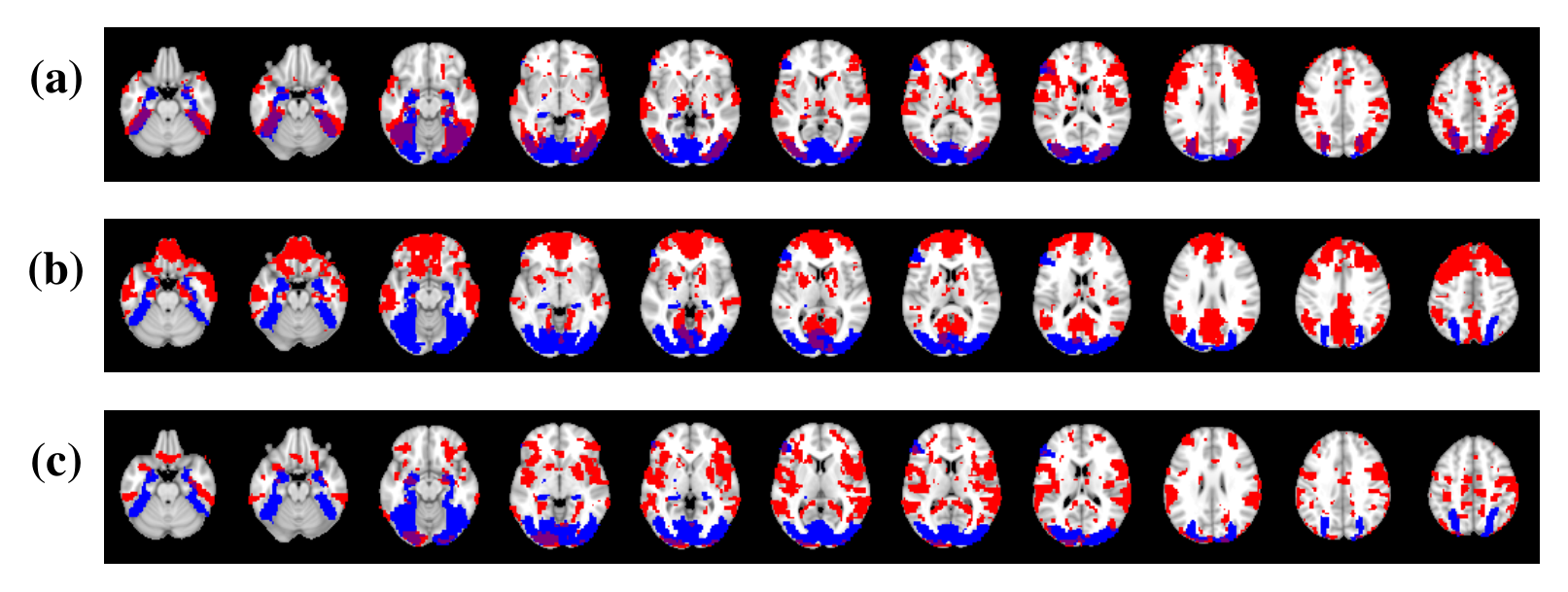}
    \caption{Representation of three output units of the restricted
             Frechet mean SOM for the \textit{unpleasant} condition in
             red, with thresholded ($p\leq0.05$) GLM $z$-score maps in blue. The output
             units have been projected in MNI-normalized space. These
             three output units are the ones that explain the largest
             amount of sample Jaccard index in that SOM. They are
             ordered by Jaccard index from panels (a) to (c),
             with the unit exhibiting the smallest Jaccard index
             in (a).
             \label{fig:unpleasant}}
\end{figure}

\sub{Visualization of Group-level SOMs}\label{sec:visualization}
In each of the three conditions, we have represented the
subject-specific restricted Frechet mean in order to produce robust visual
summaries of the different output units in each condition. This was
conducted by identifying the output units that explained the largest
amount of `variance' in the data, in terms of \textit{sample Jaccard index}. 
This measure quantifies how representative is a mean output unit
in terms of the overlap of this unit with the output units of all the
subjects in that group.

For each experimental condition, we identified the output
unit in the subject-specific SOMs that explained the largest amount of
Jaccard index. For each experimental condition, we have
plotted in figures \ref{fig:neutral}, \ref{fig:pleasant} and
\ref{fig:unpleasant}, the three output units that are associated with the least
amount of Jaccard index over all subjects. That is, for the
$j\tth$ condition, the sample Jaccard index of the $k\tth$ output
unit in the restricted Frechet mean for that condition was defined as follows,
\begin{equation}
   J(\Bm_{kj}) = 
   \frac{1}{n_{j}}\sum_{i=1}^{n_{j}} \bJ^{k}\lt(\bM_{j},\bX_{ij}\rt),
   \notag
\end{equation}
Here, $\bJ^{k}(\bM,\bX) = \sum_{v=1}^{V}\op{Jacc}\{S(\bx_{v}),S(\bw_{k})\}$
denotes the \textit{global} Jaccard distance between a mean SOM and a
subject-specific SOM. Moreover, the classical vector-specific Jaccard
distance is
\begin{equation}
  \op{Jacc}\{\bx,\by\} = \frac{C_{10} + C_{01}}{C_{11} +C_{01}+ C_{10}},
  \notag
\end{equation}
where $C_{01}$ is the number of elements satisfying $x_{h}=1$ and
$y_{h}=0$, for $h$ running from $1$ to the length of these vectors;
and $C_{10}$ and $C_{11}$ are defined similarly. 
The three output units minimizing the sample Jaccard index in
each of the three experimental conditions have been plotted in figures 
\ref{fig:neutral}, \ref{fig:pleasant} and \ref{fig:unpleasant}, for
the neutral, pleasant, and unpleasant conditions, respectively.
Allocation of a voxel to the particular output unit of a SOM is `hard', in the
sense, that either a voxel is included into an output unit or it is
included in a different one. Thus, in figures \ref{fig:neutral},
\ref{fig:pleasant} and \ref{fig:unpleasant}, we have provided the spatial
location of the voxels that have been assigned to particular output
units in each condition. 

From these three figures, one can observe that the neutral condition
is characterized by a distinct network of brain regions identified as
the third output unit in the neutral condition, as can be seen from
panel (c) of figure \ref{fig:neutral}. The set of regions associated
with this particular output unit may be interpreted as a visual
network, since it contains a considerable number of regions located in
the occipital lobe. This particular output unit was
not found to be present in the three units with the least Jaccard
index in either the pleasant or unpleasant
condition, as can be noted from figures \ref{fig:pleasant} and
\ref{fig:unpleasant}, respectively. 

\sub{Comparison with Standard GLM Maps}\label{sec:comparison}
The group-level SOM output units selected using the sample Jaccard index were
compared with standard general linear model (GLM) $z$-score
maps. Separate GLM analyses were conducted for each experimental
condition, using the FEAT function provided in the FSL software
suite \citep{Smith2004}. The $z$-score maps were thresholded at
$p=0.05$, for comparison purposes. These binarized $z$-score maps were
compared with the maps of the three `best' group-level output
units, selected on the basis of their sample Jaccard indices, as
described in the previous section. These thresholded GLM maps have
been overlaid in blue over the selected output units in figures
\ref{fig:neutral}, \ref{fig:pleasant} and \ref{fig:unpleasant}.

In each experimental condition, we computed the percentage overlap
between the maps obtained using these two different techniques. That
is, in each condition, we evaluated how many voxels were present in
both the thresholded $z$-score maps and each output
unit, normalized by the number of voxels included in the $z$-score
maps. These numerical comparisons are reported in table
\ref{tab:comparison}, where we have described the individual 
percentage overlap of the nine different output units, ranked with
respect to their sample Jaccard indices. The combined percentage overlap 
of the three output units exhibiting the lowest sample Jaccard indices
with the thresholded GLM $z$-score maps was $87\%$,
$64\%$, and $74\%$ for the neutral, pleasant
and unpleasant conditions, respectively. Although our proposed
SOM-based methods summarize such fMRI volumes in a non-linear
manner, these numerical comparisons show that the resulting output
units exhibit a considerable degree of agreement with standard GLM
analysis. 

\section{Discussion}\label{sec:discussion}
\sub{Advantages of Proposed Methods}
The main contribution of this paper is the construction of an
inferential framework for the comparison of group-level
SOMs. Although some previous researchers have considered various ways
of comparing SOMs \citep{Kaski:1996wr,Deng:2007cw,kirt2007method}, to
the best of our knowledge, no
authors have yet treated the problem of evaluating the statistical
significance of such group differences. In addition, observe that our
Frechean approach to statistical inference can be conducted for any 
choice of metrics. Indeed, the idea of defining a group distance
statistic, such as the Frechean $t$-test in equation (\ref{eq:t}) and
then evaluating its significance using permutation can be implemented
for any choice of distance functions.  
Therefore, this allows the specification of a rich
array of different distance functions capturing different aspects of
the SOMs under scrutiny. As illustrated in the main body of the paper,
we have shown that classical distance functions such as the
SMD can be modified in order to emphasize spatial, temporal or
spatio-temporal differences between the groups of interest. Here, the
choice of hypothesis is therefore superseded by the choice of
metrics over the space of SOMs. In particular, this inferential
framework allows to test hitherto untestable group-level hypotheses. 

Another substantial advantage of combining a Frechean approach with
the computation of subject-specific SOMs is that this bypasses the
problem of multiple testing correction. In standard mass-univariate
analyses of MRI volumes, one needs to control for the inflation of
the number of false positives introduced by performing a large number
of non-independent statistical tests. By contrast, we
are here conducting a single test, which identifies whether the
volumes of interest are different at a multivariate level, through the
comparison of two non-parametric unsupervised representations of the
original data. 
 
Finally, one should also note that the use of the \textit{restricted} Frechet
mean in our proposed framework is advantageous for several
reasons. On the one hand, the restricted Frechet mean greatly reduces the
computational cost of our overall analytical procedure. This is
especially true, because inference was drawn using permutations of the
group labels, and it is not clear whether such a large number of
permutations would have been possible, lest for the use of the
restricted Frechet mean. On the other hand, the restricted Frechet
mean has also the advantage of quasi-automatically transforming any distance
function into a proper metric that satisfies the four metric
axioms. This results in a non-negligible simplification of the
probabilistic theory needed to justify our inferential approach. Indeed,
most of the asymptotic results, which have been previously established
relies on the postulate that the distance function of interest is a
proper metric \citep{Ziezold1977,Sverdrup1981}.

\sub{Limitations of the Frechean Framework}
One can identify three substantial limitations to our proposed
Frechean inferential framework for SOMs, which are (i) a lack of
contrast maps, (ii) a reliance on permutation for statistical
inference, and (iii) the use of the \textit{restricted} Frechet mean in the
place of the unrestricted mean element in the space of all SOMs. 
We address these three limitations in turn. 

Firstly, one of the important limitations of our current method is that it does
not directly permit the production of a `group-difference SOM' representing
the difference between two group-specific Frechet means. In
particular, this implies that we cannot represent such differences 
by plotting a differential pattern of activation or connectivity, as is
commonly done using standard mass-univariate approaches. See the
statistical parametric network (SPN) approach, advocated by
\citet{Ginestet2011a} for example, when conducting functional network analyses.
From a neuroscientific perspective, this is a considerable limitation,
as it diminishes the interpretability of the results. We will consider
different ways of tackling this issue and producing group-difference
SOMs in future work. 

Secondly, our inferential framework has relied on permutation for
evaluating the statistical significance of the test statistics under
scrutiny. This was made computationally feasible, because this
choice of inferential method was used in conjunction with the
restricted Frechet mean. That is, for each permutation of the group
labels, the computation of the group-specific Frechet means was
straightforward, because the identification of the restricted Frechet
mean can be conducted by using the margins of the dissimilarity matrix
of the sample points --that is, the dissimilarity matrix of all the
subject-specific SOMs. Hence, the cost of calculating the group means
at each permutation was small, and the full inference could be drawn
within a couple of hours on a standard desktop computer. Such a level
of computational efficiency may not have been achieved if we had
attempted to derive the unrestricted Frechet mean, as described in
equation (\ref{eq:frechet mean}), which would have necessitated to
perform a minimization over the space of all possible SOMs. 

However, although the use of the restricted Frechet mean was
advantageous from a computational perspective, this particular
methodological choice has also its limitations. Indeed, the use of
the restricted Frechet mean in the place of the unrestricted mean
results in a loss of the classical benefits usually associated with
computing an average of real numbers. In decision-theoretic parlance
and when considering real-valued random variables, the arithmetic mean
is the quantity that minimizes the squared error loss (SEL) \citep[see][for
an introduction to decision theory]{Berger1980}. The restricted
version of the arithmetic mean for real-valued random variables would
also minimize the restricted SEL. However, the restricted arithmetic mean would
necessarily achieve a sample variance greater or equal to the one of
the unrestricted frechet mean. Note, however, that the problems
associated with the utilization of the restricted Frechet mean are also
mitigated by the fact that computing this quantity quasi-automatically
makes the space of interest a metric space, regardless of the
particular choice of distance function.

We have here used the sample Jaccard index for selecting
  the most `relevant' output units in the group-level SOMs. It
  certainly does not follow from such a selection criterion that these
  output units are of greater physiological relevance. This criterion
  is entirely statistical, and the resulting interpretation of these
  output units should remain statistical. In practice, it is advisable
  to visualize the entire set of output units obtained after this
  type of SOM analysis, in order to identify relevant physiological
  differences based on prior neuroanatomical knowledge.

\sub{Possible Extensions of these Methods}
Our proposed Frechean inferential framework could be extended in a
range of different directions. One of the most natural extensions of
this method would be to devise an $F$-test, which would generalize the
aforementioned two-sample $t_{F}$-statistic. A Frechet $F$-statistic may take the
following form. Let a data set of the form $\bM_{ij}\in (\cM,d)$, where
$i=1,\ldots,n_{j}$ labels the objects in the $j\tth$ group with
$j=1,\ldots,J$. By analogy with the classical real-valued setting, the
$F$-statistic can be defined as the ratio of the between-group to
within-group variances, $F_{F} = \op{SS}_{1}/\op{SS}_{0}$,
where these quantities are here defined with respect to the Frechet
moments, such that $\op{SS}_{1} = (J-1)^{-1}\sum_{j=1}^{J}
n_{j}d(\overline{\bM}_{j},\overline{\overline{\bM}})^{2}$, and 
$\op{SS}_{0} = (N-J)^{-1}\sum_{j=1}^{J}\sum_{i=1}^{n_{j}}
d(\bM_{ij},\overline{\bM}_{j})^{2}$, using standard notation for the Frechet
sample group means, $\overline{\bM}_{j}$, and grand mean,
$\overline{\overline{\bM}}$. One can then 
test for the null hypothesis that $H_{0}:\sig^{2}_{1} =
\sig^{2}_{0}$, where $\sig^{2}_{1}$ and $\sig^{2}_{0}$ are the
theoretical equivalents of $\op{SS}_{1}$ and $\op{SS}_{0}$,
respectively. Statistical inference can, again, be conducted using
permutation of the group labels.

In addition, the analytical strategy that we have here described could also
be improved through the use of different types of SOM algorithm. In
the present paper, we have made use of the batch SOM
algorithm. However, several other alternatives to the traditional
sequential SOM algorithm have been proposed in the literature. In
particular, \citet{vesanto2000} have showed that the SOMs obtained
when using the batch SOM algorithm with an initialization of the maps
based on the eigenvectors of the input data can produce more robust
results. Since every SOM is computed independently for each subject,
such an improvement of the existing batch SOM algorithm could easily
be incorporated in our proposed inferential framework. 

One of the outstanding questions that is implicitly raised in
  this paper is the possibility of separately weighting the individual
  contributions of the spatial and temporal properties contributing to
  the overall SOM difference. Such a question is likely to be arduous
  to answer, however, since the temporal and spatial properties of the fMRI
  volumes of interest necessarily live in distinct abstract
  spaces. On one hand, the temporal differences in T-SMD were
  quantified using a Euclidean distance in a $T$-dimensional vector
  space; whereas, on the other hand, the spatial differences in S-SMD
  were quantified using the Hamming distance on binary vectors of
  varying sizes. It is unclear whether the magnitude of the distances
  in these different metric spaces could be normalized in order to
  ensure a modicum of comparability.

\section{Conclusions}\label{sec:conclusion}
In this paper, we have described a formal framework for drawing
group-level inference between unsupervised multivariate summaries of
fMRI data. Our proposed approach proceeds by computing
subject-specific SOMs, and computing the sample Frechet
mean in each group of subjects. Despite the unwieldy nature of the
space of all possible SOMs, this can be done efficiently by
identifying the restricted Frechet mean. Statistical inference on the
difference between the group restricted Frechet means can be conducted
using permutation on the group labels. This framework can be
implemented for any choice of metrics quantifying the difference
between pairs of SOMs. As such, the specification of a
particular distance function is equivalent to the choice of a particular
hypothesis test. Different researchers may therefore be
interested in evaluating different metrics, which capture different
aspects of the SOMs.

We have hence described and evaluated several types of distance
functions for SOMs based on fMRI data. In particular, we have considered
variants of the classic SMD function, which has previously been used
to compare pairs of SOMs. Our proposed variants distinguish between the
temporal, spatial and spatio-temporal properties of the data under
scrutiny. Our inferential framework and these metrics were tested on
both synthetic and real data, Our analysis of the simulated data
showed that the distance functions of interest were indeed capturing
the aspects of the data that they were purported to measure. 
In addition, the findings of the re-analysis of an fMRI experiment
has demonstrated the capacity of these methods to extract new
information from existing data sets. In this paradigm, the differences of 
the restricted mean SOMs in the pleasant and unpleasant conditions
were found to be smaller than the differences between the mean SOMs in
any of these two conditions with respect to the one in the neutral
condition. 

Taken together, the analyses of these synthetic and real data sets
have underlined the robustness and potential usefulness of these methods.
It is hoped that this type of global inferential perspective on
neuroimaging data will inspire other neuroscientists to follow this
research avenue. One could imagine a range of other subject-specific
abstract-valued random variables that could be suitably analyzed using this
type of Frechet inferential framework. In fact, the very use of
mass-univariate approaches in the context of neuroimaging could be
superseded by a more global perspective, where a single statistical
test is conducted, thereby bypassing the need for exacting multiple testing
penalties.

\appendix 
\section{Appendices}
\subsection{A. From Distance Functions to Metrics}\label{sec:A}
Let $d$ be a distance function on a \textit{finite} space of SOMs,
$\bLa=\{\bM_{1},\ldots,\bM_{n}\}$, which satisfies the
positivity, coincidence and symmetry axioms. In order to transform the
distance function $d$ into a proper metric $\widetilde{d}$ satisfying
the triangle inequality, we need to construct a saturated graph $G=(V,E)$
representing the topology of $\bLa$. The vertex set of $G$ is defined
as $V(G)= \bLa$. Its edge set is composed of all the possible links
between the elements of $\bLa$. That is, $G$ is a saturated graph, in
the sense that it contains the maximal number of edges. Each of these
edges is denoted by $\bM_{i}\bM_{j}\in E(G)$, for any $0\leq i\neq
j\leq n$. 

A \ti{path} in $G$ is a non-empty subgraph $P\subseteq G$ of
the form $V(P)=\{\bM_{0},\ldots,\bM_{k}\}$ and
$E(P)=\{\bM_{0}\bM_{1},\ldots,\bM_{k-1}\bM_{k}\}$, where the $\bM_{i}$'s
are all \textit{distinct}. Following an idea proposed by
\citet{ThomasEiter:1997ux}, it is now possible to construct a new
distance function, denoted $\widetilde{d}$, defined as the set of 
\textit{shortest paths} in $\bLa$, such that for any
$\bM,\bM\pri\in\bLa$, we have
\begin{equation}
  \notag
  \widetilde{d}(\bM,\bM\pri) =
  \min_{P\in\cP(\bM,\bM\pri)} \sum_{\bM_{i}\bM_{j}\in E(P)} d(\bM_{i},\bM_{j}),
\end{equation}
where $\cP(\bM,\bM\pri)$ is the set of all paths in $G$ between
$\bM$ and $\bM\pri$. 
By construction, it immediately follows that $\widetilde{d}$ satisfies
the triangle inequality. Therefore, $(\bLa,d)$ forms a proper metric
space. 

\subsection{B. Choice of SOM Dimensions}\label{sec:B}
A supplemental set of simulations was conducted in order to
investigate the effect of the choice of SOM dimensions on group-level
statistical inference. We assessed the effect of rectangular SOMs, as
well as the effect of increasing the dimensions of these maps. The
synthetic data used for these simulations followed the design
described in the section entitled Synthetic Data Simulations, based on
the three different scenarios, and using the three types of SMD
functions described in this paper, and setting $\op{SNR}=1$. These
results are reported in table \ref{tab:size}.

The results of these simulations were consistent with the ones
described in our first analysis of these synthetic data. In particular,
for any choice of SOM dimensions, we obtained strong corroborations of
the previous findings. Under both SC1 and SC2, the T-SMD tended to
outperform its counterparts for any choice of SOM dimensions. As
before, S-SMD performed poorly throughout these simulations,
irrespective of the choice of SOM dimensions. Finally, the
ST-SMD function exhibited good performance on all scenarios, and
outperformed the T-SMD under SC3, although ST-SMD did not reach
significance level for this particular scenario. 
\begin{table}[t]
\centering
\begin{tabular}{ c | c | c | c | c }
\textit{Scenarios} & \textit{SOM Dimensions} & T-SMD & S-SMD & ST-SMD\\
\hline
SC1 (Spatio-temporal)
       & $10 \times 10$ & $0\pm0$ & $0.626\pm0.236$ & $0.002\pm0.064$ \\       
       & $5 \times 5$ & $0\pm0$ & $0.498\pm0.256$ &  $0.001\pm0.031$ \\       
       & $4 \times 6$ & $0\pm0$ & $0.516\pm0.285$ & $0.001\pm0.012$ \\       
       & $6 \times 8$ & $0\pm0$ & $0.464\pm0.279$ & $0.001\pm0.078$ \\      
\hline
SC2 (Temporal)  & $10 \times 10$ & $0\pm0$ & $0.612\pm0.350$ & $0\pm0$ \\       
       & $5 \times 5$ & $0\pm0$ & $0.557\pm0.298$ &  $0.011\pm0.014$ \\       
       & $4 \times 6$ & $0\pm0$ & $0.474\pm0.288$ & $0.002\pm0.012$ \\       
       & $6 \times 8$ & $0\pm0.006$ & $0.487\pm0.269$ & $0.003\pm0.097$ \\       
\hline
SC3 (Spatial)  & $10 \times 10$ & $0.505\pm0.296$ & $0.523\pm0.282$ & $0.180\pm0.171$ \\       
       & $5 \times 5$ & $0.519\pm0.206$ & $0.504\pm0.279$ &  $0.149\pm0.144$ \\       
       & $4 \times 6$ & $0.482\pm0.272$ & $0.559\pm0.268$ & $0.108\pm0.162$ \\       
       & $6 \times 8$ & $0.482\pm0.300$ & $0.451\pm0.281$ & $0.149\pm0.103$ \\       
\hline
\end{tabular}
\caption{Simulation results with varying SOM dimensions summarized as
         mean significance levels and standard deviations of these distributions, based on synthetic data
         with 100 simulations in every cell and $\op{SNR}=1$. These results are reported for the three
         scenarios described in figure \ref{fig:sim}, which are denoted by
         SC1, SC2 and SC3, for three different levels of SNR, and for
         the three different distance functions under scrutiny, denoted by
         T-SMD, S-SMD and ST-SMD, which stand for the temporal SMD,
         spatial SMD, and spatio-temporal SMD, respectively. These
         results are consistent with the ones of table \ref{tab:sim}.
  \label{tab:size}}
\end{table}

\small
\addcontentsline{toc}{section}{References}
\bibliographystyle{/home/cgineste/ref/style/oupced3}
\bibliography{/home/cgineste/ref/bibtex/Statistics,%
              /home/cgineste/ref/bibtex/Neuroscience,%
              soms_frechet}
\end{document}